\documentclass[epj]{webofc}
\usepackage[utf8]{inputenc}
\usepackage[varg]{txfonts}   % Web of Conferences font
\usepackage{booktabs}
\usepackage{xcolor}
\definecolor{darkred}{rgb}{0.4,0.0,0.0}
\definecolor{darkgreen}{rgb}{0.0,0.4,0.0}
\definecolor{darkblue}{rgb}{0.0,0.0,0.4}
\usepackage[bookmarks,linktocpage,colorlinks,
    linkcolor = darkred,
    urlcolor  = darkblue,
    citecolor = darkgreen]{hyperref}
\wocname{EPJ Web of Conferences}
\woctitle{Lattice2017}
\def\Eq#1{Eq.~(\ref{#1})}

\begin{document}
%%%%%%%%%%%%%%%%%%%%%%%%%%%%%%%%%%%%%%%%%%%%%%%%%%%%%%%%%%%%%%%%%%%%%%%%%%%%%
%
\selectlanguage{english}
%----------------------------------------------------------------------------
\title{%
Looking behind the Standard Model with lattice gauge theory 
}
%----------------------------------------------------------------------------
\author{%
\firstname{Benjamin} \lastname{Svetitsky}\inst{1}\fnsep
\thanks{
Supported by the Israel Science Foundation under Grant No.~449/13.\\
Plenary lecture given at Lattice 2017, the 35th International Symposium on Lattice Field Theory, Granada, Spain, 18--24 June 2017.}
\fnsep\thanks{\email{bqs@julian.tau.ac.il}
}
% etc.
}
%----------------------------------------------------------------------------
\institute{%
School of Physics and Astronomy, Tel Aviv University, 6997800 Tel Aviv, Israel}
%----------------------------------------------------------------------------
\abstract{%
Models for what may lie behind the Standard Model often require non-perturbative calculations in strongly coupled field theory. This creates opportunities for lattice methods, to obtain quantities of phenomenological interest as well as to address fundamental dynamical questions. I  survey recent work in this area.
}
%----------------------------------------------------------------------------
\maketitle
%----------------------------------------------------------------------------
%%%%%%%%%%%%%%%%%%%%%%%%%%%%%%%%%%%%%%%%%%%%%%%%%%%%%%%%%%%%%%%%%%%%%%%%%%%
\section{Introduction}\label{intro}
%%%%%%%%%%%%%%%%%%%%%%%%%%%%%%%%%%%%%%%%%%%%%%%%%%%%%%%%%%%%%%%%%%%%%%%%%%%

What does the B in BSM stand for?  Beyond, Behind, or what? 

We can begin a list of what might lie {\em beyond\/} the Standard Model.
First, of course, anything new to be discovered in experiment is beyond the SM by definition---new particles (none yet), or flavor physics anomalies (maybe~\cite{Oyanguren,Becirevic}).
In the realm of theory, there are phenomena that find no explanation in the SM: the matter--antimatter asymmetry, dark matter, dark energy, the fields responsible for inflation, the incorporation of gravity.
All of these are {\em beyond\/} the Standard Model simply because they're not {\em in\/} the Stadard Model.

The SM itself, however, poses puzzles of its own.
We ask what lies {\em behind\/} the SM that makes it what it is.
Why the $\textrm{SU}(3)\times\textrm{SU}(2)\times\textrm{U}(1)$ gauge structure?
Why three families?
Where do the quark and lepton masses and mixing angles come from, and what creates their hierarchical structure?
Ditto for neutrino masses and angles, which, taken together with the quarks and leptons, make the hierarchies even more mysterious.
And the Higgs boson.

The Higgs boson takes some explaining, and it's not just another number to explain.
It has been a {\em b\^ete noire\/} for theorists for forty years.
Originally predicted to emerge as the radial mode of a fundamental scalar field, its mass, whatever it is, is an unnatural number.
This is embodied in the cartoonish equation of mass renormalization,
\begin{equation}
m_H^2=m_0^2+\textit{const}\cdot M_{\textrm{Planck}}^2\,,
\label{unnatural}
\end{equation}
where $m_0^2$ is to be tuned to balance (125 GeV)$^2$ on the left against ($10^{19}$ GeV)$^2$ on the right.

The origin of \Eq{unnatural} is the quadratic divergence that gives an additive mass renormalization in scalar field theory.
This divergence is reduced to a logarithmic divergence in supersymmetric theories, where scalars are paired with fermions whose mass is protected by chiral symmetry.
Then the tuning of parameters in the Lagrangian is not so frightening.
Signs of supersymmetry, however, are remarkably absent at the LHC, so I'll set this subject aside for this lecture.\footnote{A better excuse is that the field of lattice supersymmetry has been rather quiet in the past year.
On the other hand, several talks on the subject have been given in the parallel sessions at this conference.}

In this lecture I will discuss recent work on some approaches to a composite Higgs boson.
I refrain from starting each story at its beginning because there are excellent and lengthy reviews in the literature \cite{DeGrand:2015zxa,Nogradi:2016qek}, including Claudio Pica's plenary lecture at last year's Lattice conference \cite{Pica:2017gcb}.
Thus I will limit the scope of the lecture and (most) citations to work that has appeared in the last year.
The bulk of this work has been concerned with theories descended from technicolor ideas, in which the mechanism of {\em walking technicolor\/} is supposed to produce a light {\em dilatonic Higgs}.
An alternative is to produce a composite Higgs as a pseudo-Goldstone boson, a mechanism known as ``composite Higgs'' in the narrow sense.
This, also, is the subject of recent and current work.

%%%%%%%%%%%%%%%%%%%%%%%%%%%%%%%%%%%%%%%%%%%%%%%%%%%%%%%%%%%%%%%%%%%%%%%%%%%
\section{Ultraviolet completions}
%%%%%%%%%%%%%%%%%%%%%%%%%%%%%%%%%%%%%%%%%%%%%%%%%%%%%%%%%%%%%%%%%%%%%%%%%%%

Taking the SM to be an effective theory limited to low energies, one seeks an ultraviolet theory from which it can emerge.
In particular, while the rest of the SM fields might be fundamental, the Higgs field should emerge as some kind of composite.
Attempts at {\em UV completions\/} have two features.
Fundamentally, and at high energies, they are based on an asymptotically free gauge theory with spin-$1/2$ matter, so that fundamental renormalizations are only logarithmically divergent.
Then the Higgs particle, or maybe the entire Higgs multiplet of the SM, emerges as a {\em composite\/} field ruled by a low-energy effective theory.
Integrals in this theory are naturally cut off at the energy scale $\Lambda$ of compositeness, where the theory must be replaced by the UV completion.
Calculation in the effective theory would renormalize the Higgs mass once more according to \Eq{unnatural}, with $\Lambda$ replacing $M_{\textrm{Planck}}$; if $\Lambda$ is not too large then $m_H=125$~GeV could be quite natural.

The original idea of {\em technicolor\/} (see below) took $\Lambda$ to be around the Higgs expectation value $v\simeq 245$~GeV, making the Higgs mass plausible, but one would expect a composite theory to have other composite particles with masses on the order of $\Lambda$.
Today the LHC, through the absence of discoveries, has bounded the compositeness scale by $\Lambda\!\gtrsim\!5$~TeV so \Eq{unnatural} still has a bite.
Any composite model must explain why there is a scalar particle with $m_H\ll\Lambda$ while the rest of the spectrum is at $\sim\Lambda$ or higher.

A distinct problem is how a theory with scale $\Lambda$ can produce the correct value of $v$, which is still fixed at the weak scale.
In technicolor, $v$ is determined by $f_\pi$, the decay constant of the technipions.
So something has to detach $f_\pi$ from $\Lambda$ and from the spectrum of technihadrons---more on this in Sec.~\ref{sec:dilaton}.

Generally speaking, composite states emerge from strong dynamics.
The need to understand strong dynamics in the UV theory is what brings lattice gauge theory into the picture.
When one casts an eye over the published lattice work, one perceives a rule of procedure:
{\em Don't bite off too much.}
This means that in a given project, one focuses on a specific phenomenon, in some class of model, that explains something but not everything.
I think this coincides with common practice in modern phenomenology, whereby one tries to connect sub-LHC physics (e.g., $m_H=125$~GeV) to a super-LHC model (with $\Lambda\!\gtrsim\!5$~TeV)
$\ldots$ without trying to explain everything at once.
I believe that this is quite different from the ideals of phenomenology in the 1970's and 80's.

By the way, most composite-Higgs phenomenology works with effective actions for the Higgs field and other scalars, for instance nonlinear sigma models, without even asking about true ultraviolet completions.
This just reflects the difficulty of calculating things in strongly coupled theories, which, as I said before, invites lattice gauge theorists to take a hand.
Checking out an ultraviolet completion of an effective model is like sliding under a car to see what makes it work---not a bad metaphor for lattice gauge theory as a whole.
So now we have another meaning of BSM:  {\em Beneath\/} the Standard Model.

%%%%%%%%%%%%%%%%%%%%%%%%%%%%%%%%%%%%%%%%%%%%%%%%%%%%%%%%%%%%%%%%%%%%%%%%%%%
\section{Technicolor}
%%%%%%%%%%%%%%%%%%%%%%%%%%%%%%%%%%%%%%%%%%%%%%%%%%%%%%%%%%%%%%%%%%%%%%%%%%%

Whatever the UV theory is that produces the Higgs boson, it's not QCD.
The problem is the hierarchy one needs to produce between the Higgs mass $m_H$ and the compositeness scale $\Lambda$.
QCD's lightest scalar is known to nuclear physicists as $\sigma$ and to the Particle Data Group as $f_0$.
Its mass lies between 400 and~500~MeV, to be compared to the QCD scale as given by the lightest vector, $m_\rho=770$~MeV.
This isn't much of a hierarchy.
Moreover, because of the decay $\sigma\to\pi\pi$ into light pseudo-Goldstone bosons, the $\sigma$ is very broad, $\Gamma_\sigma=400$--700~MeV.
The Higgs width, by contrast, is already bounded experimentally by $\Gamma_H\!\lesssim\!2$--5~GeV (depending on the measurement) so it had better not decay strongly to lighter bound states of the UV theory.

The originators of technicolor didn't know yet that $\Lambda$ would be so large, so their prototype was a copy of two-flavor, massless QCD with $\Lambda$ and $f_\pi$ scaled up to the order of $v\simeq245$~GeV.
Let's follow their logic.
Copying out of QCD's notebook, we know that the $\textrm{SU}(2)_L\times\textrm{SU}(2)_R$ chiral symmetry is spontaneously broken to SU(2)$_V$, producing three Nambu--Goldstone bosons named $\pi^\pm,\pi^0$.
The broken generators include three generators of the $\textrm{SU}(2)_L\times\textrm{U}(1)$ gauge group of the electroweak interaction and thus we have a Higgs phenomenon without a fundamental scalar field to bring it about.
The three NG bosons are swallowed by the (fundamental) $W^\pm$ and $Z$ gauge bosons and give them mass.
Three NG bosons, with nothing left over---so there is no  Higgs boson at all.

Technicolor sank beneath the waves as its compositeness scale $\Lambda$ was pushed ever higher by experiment.
It appeared to go down for the third time when the Higgs was found, just five years ago.
There is, however, a mechanism that may produce a scalar much lighter than $\Lambda$, and of course it is based on a departure from the QCD paradigm, obtained most simply by raising the number of flavors.
This introduces an approximate scale invariance, which results in a light scalar boson.
 
\begin{figure}[hbt]
  \centering
  \sidecaption
  \includegraphics[width=7cm,clip]{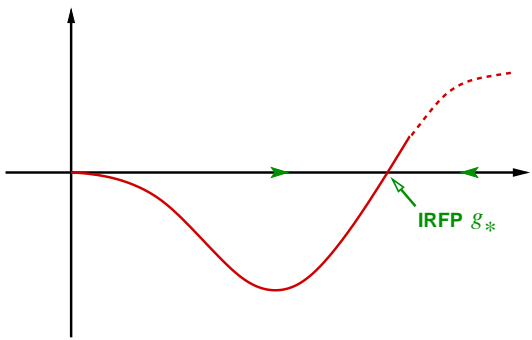}
  \caption{The beta function for a gauge theory in the conformal window. The green arrows indicate infrared flow.}
  \label{fig:beta1}
\end{figure}

Whence this scale invariance:  As everybody knows, the $\beta$ function of QCD with a physical number of flavors is strictly negative, so as the length scale grows, the theory flows from asymptotic freedom in the UV to strong coupling, confinement, and a chiral condensate in the IR.
For a large number of flavors, however, the $\beta$ function crosses the axis and becomes positive (see Fig.~\ref{fig:beta1}), creating an IR-stable fixed point at some finite coupling $g_*$.
%%%%%%%%%%%%%%%%%%%%%%%%%%%%%%%%%%%%%%%%%%%%%%%%%%%%%%%%%%%%%%%%%%%%%%%%%%%
%%%%%%%%%%%%%%%%%%%%%%%%%%%%%%%%%%%%%%%%%%%%%%%%%%%%%%%%%%%%%%%%%%%%%%%%%%%
%%%%%%%%%%%%%%%%%%%%%%%%%%%%%%%%%%%%%%%%%%%%%%%%%%%%%%%%%%%%%%%%%%%%%%%%%%%
\begin{figure}[hbt]
  \centering
  \sidecaption
  \includegraphics[width=7cm,clip]{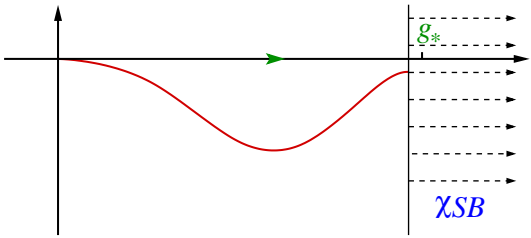}
  \caption{The beta function for a gauge theory just below the sill of the conformal window.
  The function is undefined once the chiral condensate forms, before the IRFP at $g_*$ is reached.}
  \label{fig:beta2}% Give a unique label
\end{figure}
%%%%%%%%%%%%%%%%%%%%%%%%%%%%%%%%%%%%%%%%%%%%%%%%%%%%%%%%%%%%%%%%%%%%%%%%%%%
This happens when the number of flavors $N_f$ is greater than some threshhold value $N_f^*$ (we keep $N_f<16\frac12$ so as not to lose asymptotic freedom).
When the coupling flows out of asymptotic freedom it gets stuck at the IR fixed point; at large distances, then, the theory is scale-invariant and hence conformally invariant, so that there is no confinement scale, no chiral condensate, no particle masses, and indeed no particles in the spectrum.
Such values of $N_f$ are said to be in the {\em conformal window\/} and $N_f^*$ is the sill of the window.

Just below the sill, that is, when $N_f$ is just below $N_f^*$, the $\beta$ function doesn't quite cross the axis before the coupling reaches a critical value that induces a chiral condensate like that of QCD (see Fig.~\ref{fig:beta2}).
In this scenario, just before the condensation, the running of the theory slows to a walk and the theory is nearly scale invariant over a wide range of scales.
This is called {\em walking technicolor}.\footnote{The original motivation for walking was to induce physically correct fermion masses while suppressing flavor-changing neutral currents in extended technicolor theories.
I won't go down this path here.
{\em ``Don't bite off too much''} can mean {\em one thing at a time.}}
If there is an approximate dilatation invariance, then it can have a pseudo-Goldstone boson called the {\em dilaton,} and this is supposed to be the anomalously light scalar that we see as the Higgs boson.
Its mass is protected from quadratic divergences in the UV because it is composite.
Moreover, it couples to SM particles via the trace $T^{\ \mu}_\mu$ of the stress--energy tensor (see Sec.~\ref{sec:dilaton}), so its couplings are proportional to particle masses just like the conventional (and the experimental!) Higgs.

I'll say more about this scenario as I review recent calculations.
There are two main issues pursued by lattice efforts:
\begin{enumerate}
\item For a given gauge group and fermion representation, where is the sill $N_f^*$?
\item Is there a light scalar in theories below the sill?
\end{enumerate}
The answers today seem to be
\begin{enumerate}
\item It's hard to nail this down.
It might not matter too much, however, as long as you keep $N_f$ safely below the sill in your search for a light Higgs.
\item Yes.
\end{enumerate}

%%%%%%%%%%%%%%%%%%%%%%%%%%%%%%%%%%%%%%%%%%%%%%%%%%%%%%%%%%%%%%%%%%%%%%%%%%%
\section{Finding the sill: SU(3) with 12 flavors}
%%%%%%%%%%%%%%%%%%%%%%%%%%%%%%%%%%%%%%%%%%%%%%%%%%%%%%%%%%%%%%%%%%%%%%%%%%%

Let's focus on the SU(3) gauge theory with $N_f$ flavors of color-triplet fermions---generalized QCD.
The two-loop $\beta$ function crosses zero if $N_f>8.05$, so this is a first guess at $N_f^*$;
it is clear, however, that for $N_f$ just greater than 8.05 the would-be fixed point $g_*$ is at very strong coupling, so there must be a condensate induced before the coupling runs that far.
Hence if you want a real fixed point then $N_f$ should be well above 8.05.
%$N_f^*$ should be well above the two-loop value.
In lattice simulations, a lot of work has been done with $N_f=4n$ because it is convenient for staggered fermions, thus leading to a focus on $N_f=8$ and~12.

Today we have a long-running controversy over whether the $N_f=12$ theory is confining or conformal.
The question has been addressed with many methods, studying in turn the scaling of the particle spectrum as the quark mass is taken to zero; the possible disappearance of the finite-temperature phase transition; the scaling of Dirac eigenvalues.
All these try to distinguish between conventional QCD-like behavior and novel conformal behavior.
The predominant conclusion is that the theory is conformal.

There is a danger, however, in applying methods developed for QCD to this issue.
Basically, one is trying to tell {\em slow\/} running---meaning a walking towards a QCD-like condensate---from {\em no\/} running, that is, stopping at an IR fixed point.
If there is fixed-point physics at an IR scale $L$---say, the size of the lattice---then it will typically involve a strong (i.e., nonperturbative) coupling.
Whether the running is either slow or (almost) stopped, the coupling at the UV cutoff---the lattice spacing $a$---will be strong as well.
This makes it very difficult to take a continuum limit.

The renormalization group was created for this very purpose.
One compares two scales $L_1$ and $L_2$ to derive a $\beta$ function, which then relates very small to very large scales.
One can compare this $\beta$ function to the conjectured forms on either side of the conformal sill, which was how the problem was formulated in the first place.
Taking a continuum limit, though, is still a subtle matter, and differs considerably from how one does it for QCD.
On a given lattice, there will be no dynamical scale $\Lambda$ between $1/a$ and $1/L$, so physical quantities must be functions of $a/L$ alone.
Thus the continuum extrapolation $a\to0$ is equivalent to a large-volume limit $L\to\infty.$

The latest results on the $N_f=12$ theory thus come from RG studies employing the gradient flow to define the running coupling \cite{Lin:2015zpa,Fodor:2016zil,Hasenfratz:2016dou}.
In comparing different calculations, one should keep the following in mind:
\begin{itemize}
\item The existence or nonexistence of a fixed point is a universal issue.
The answer should be independent of the scheme for defining the running coupling.
The same goes for critical indices/anomalous dimensions.
\item The location $g_*$ of the fixed point is not universal.
Neither is the shape of the $\beta$ function.
These can vary with the scheme.
For instance, older calculations used the Schr\"odinger functional while the current work uses (largely) the gradient flow.
Even among the latest calculations, however, there is a parameter $c$ in defining the coupling.
On a lattice of linear size $L$, one defines the running coupling
\begin{equation}
g_{\rm{GF}}^2=\frac{128\pi^2}{3(N^2-1)}\left\langle{t^2E(t)}\right\rangle
\end{equation}
through integrating the flow equations to a certain $t=t_0$.
This defines a distance scale $\sqrt{t_0}$ that has to be taken proportional to $L$ so that there will be only one scale in the problem, leading thus to the requirement
\begin{equation}
\sqrt{8t_0}=cL.
\end{equation}
Changing $c$ may be convenient but it changes the renormalization scheme and thus the $\beta$ function.
\item Even so, the $\beta$ function with its supposed fixed point should not depend on the discretization or improvement of the lattice action, since one always extrapolates to the continuum.
Likewise it should be robust against variations in the discretization of the flow equations and of the definition of the flow energy $E$.
Finally, while different calculations may use different scale factors $s\equiv L_1/L_2$ in a discrete rescaling, it is not difficult to reduce them to a common value of $s$ and to derive the conventional $\beta$ function as $s\to1$.
\end{itemize}

%%%%%%%%%%%%%%%%%%%%%%%%%%%%%%%%%%%%%%%%%%%%%%%%%%%%%%%%%%%%%%%%%%%%%%%%%%%
\begin{figure}[hbt]
  \centering
  %\sidecaption
  \includegraphics[width=9cm,clip]{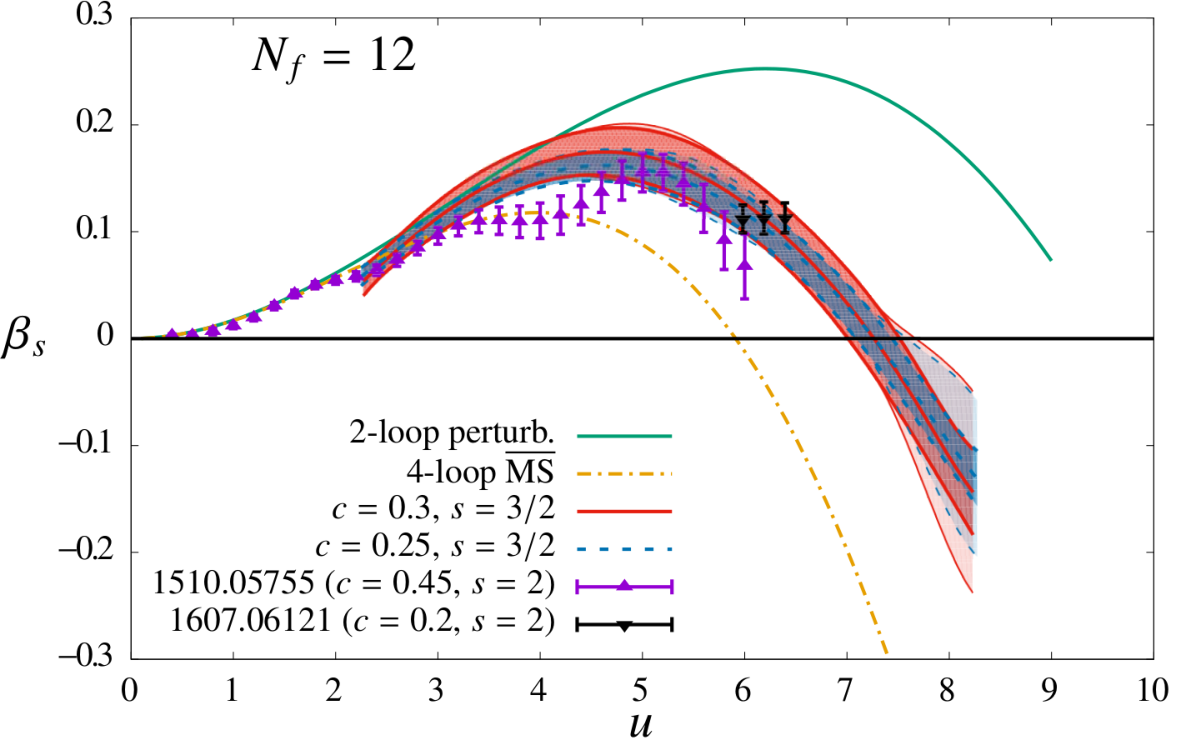}
  \caption{The discrete beta function of SU(3) gauge theory with 12 flavors from Ref.~\cite{Hasenfratz:2016dou}, showing as well data from Refs.~\cite{Lin:2015zpa,Fodor:2016zil}.
  The abscissa is $u=g^2$ and the various curves and data have been recalculated for a common scale factor $s=3/2$ for purposes of comparison.}
  \label{fig:anna}
\end{figure}
%%%%%%%%%%%%%%%%%%%%%%%%%%%%%%%%%%%%%%%%%%%%%%%%%%%%%%%%%%%%%%%%%%%%%%%%%%%
The results of the three papers cited are summarized in Fig.~\ref{fig:anna}, taken from Hasenfratz and Schaich (HS) \cite{Hasenfratz:2016dou}.
The $\beta$ function calculated by HS (the red and grey bands) is consistent with the earlier two papers (purple and black points), but HS are able to go to stronger couplings through the use of an nHYP fermion action and a two-term gauge action.
This consistency is a bit curious, in light of the above remarks; indeed, none of the curves or data sets plotted in Fig.~\ref{fig:anna} need agree with each other at the quantitative level, since they are all scheme-dependent.
In any case, HS find clear evidence for an IR fixed point.

There is new evidence on the other side, too.
Fodor {\em et al.}, the authors of Ref.~\cite{Fodor:2016zil} and of the black points in Fig.~\ref{fig:anna}, have extended their calculation to stronger couplings as well and they {\em exclude\/} a zero of the $\beta$ function in the region of the HS zero \cite{Fodor}.
To compare, one should look carefully at the respective analyses of volume dependence and the choice of $c$.

Anna Hasenfratz has noted that an IR fixed point could be destroyed by perturbations that break the global ${\rm SU}(N_f)\times{\rm SU}(N_f)$ symmetry of the continuum theory and has provided examples from the theory of critical phenomena \cite{Anna}.
This could spell trouble for any analyses---including all the above---that are based on staggered fermions with their inherent taste breaking.
At a naive level, broken taste symmetry would mean that the theory under study has fewer effective massless flavors, which would bias the result towards confinement rather than conformality \cite{DeGrand:2010na}.

It is important to note that T.-W.~Chiu \cite{Chiu:2016uui,Chiu:2017kza} has carried out a gradient flow RG calculation for the theory with $N_f=10$, defined with domain wall fermions.
He has concluded that there is indeed an IR fixed point.
If the $N_f=10$ theory is conformal, it would be strange indeed if the $N_f=12$ theory confines.

%%%%%%%%%%%%%%%%%%%%%%%%%%%%%%%%%%%%%%%%%%%%%%%%%%%%%%%%%%%%%%%%%%%%%%%%%%%
\section{A light scalar: SU(3) with 8 flavors}
%%%%%%%%%%%%%%%%%%%%%%%%%%%%%%%%%%%%%%%%%%%%%%%%%%%%%%%%%%%%%%%%%%%%%%%%%%%

As mentioned above, the two-loop $\beta$ function of the SU(3) gauge theory develops an IR fixed point for $N_f>8.05$, and dynamical arguments place the nonperturbative value of $N_f^*$ well above this.
The 8-flavor theory, then, should be well below the conformal window.
The spectrum of this theory has been examined recently by the LatKMI collaboration \cite{Aoki:2016wnc,Rinaldi}, following earlier (and continuing) work by the LSD collaboration \cite{Appelquist:2014zsa,Appelquist:2016viq,Fleming}.

The claim of both collaborations is that the 8-flavor theory {\em walks,} and this creates a light scalar.
This is based on the dependence of the spectrum on the fermion mass $m_f$.
First, as $m_f\to0$ one finds signs of the formation of a chiral condensate: $F_\pi$ goes to a nonzero constant while $m_\pi$ goes to zero.
Nonetheless, there is a wide range of $m_f$ where all the ``hadron'' masses---except that of the $\pi$---scale according to
\begin{equation}
M_H\sim m_f^{1/(1+\gamma)},    \label{hyper}
\end{equation}
with a common exponent $\gamma\simeq1$ for all $H$.
This {\em hyperscaling\/} is a sign of a nearby fixed point, avoided first by the nonzero $m_f$ and then by the formation of the condensate in the IR.

\begin{figure}[hbt]
  \centering
  %\sidecaption
  \includegraphics[width=9cm,clip]{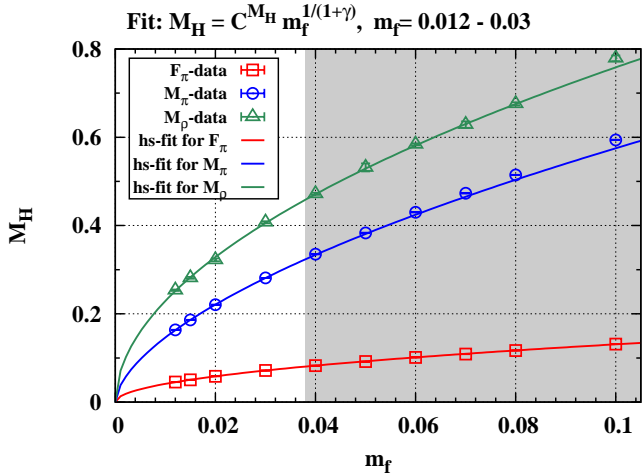}
  \caption{Power-law fits to hadronic quantities as a function of the fermion mass in the $N_f=8$ theory, from Ref.~\cite{Aoki:2016wnc}.}
  \label{fig:hyper}
\end{figure}

\begin{figure}[hbt]
  \centering
  %\sidecaption
  \includegraphics[width=9cm,clip]{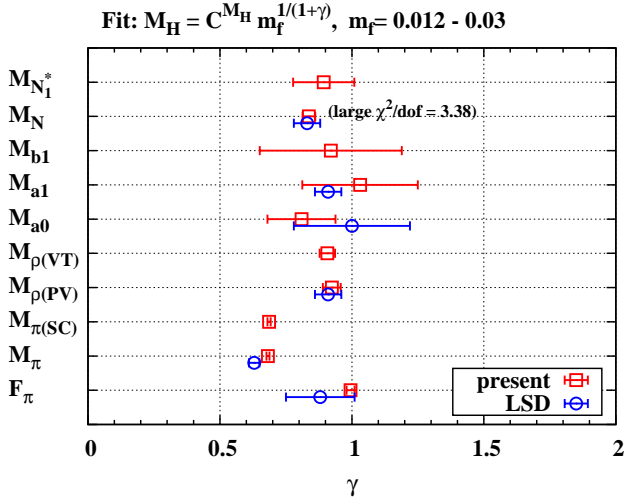}
  \caption{Scaling exponents for hadron masses and $F_\pi$, from Ref.~\cite{Aoki:2016wnc} (quoting Ref.~\cite{Appelquist:2014zsa}).}
  \label{fig:gamma}
\end{figure}

Let's look at hyperscaling first.
Figure~\ref{fig:hyper} from LatKMI shows how well power-law scaling works for masses of the $\pi$ and~$\rho$ and for $F_\pi$ at intermediate values of $m_f$.
%%%%%%%%%%%%%%%%%%%%%%%%%%%%%%%%%%%%%%%%%%%%%%%%%%%%%%%%%%%%%%%%%%%%%%%%%%%
%%%%%%%%%%%%%%%%%%%%%%%%%%%%%%%%%%%%%%%%%%%%%%%%%%%%%%%%%%%%%%%%%%%%%%%%%%%
What is not evident is that the exponent for $m_\pi$ is different from the other two.
This may be seen in Fig.~\ref{fig:gamma}, which includes as well the exponents calculated by LSD in earlier work using domain-wall fermions~\cite{Appelquist:2014zsa}: the pion and its taste partner do not scale with the other masses or with $F_\pi$, all of which in turn do scale with a common exponent.
%%%%%%%%%%%%%%%%%%%%%%%%%%%%%%%%%%%%%%%%%%%%%%%%%%%%%%%%%%%%%%%%%%%%%%%%%%%
%%%%%%%%%%%%%%%%%%%%%%%%%%%%%%%%%%%%%%%%%%%%%%%%%%%%%%%%%%%%%%%%%%%%%%%%%%%

\begin{figure}[hbt]
  \centering
  %\sidecaption
  \includegraphics[width=9cm,clip]{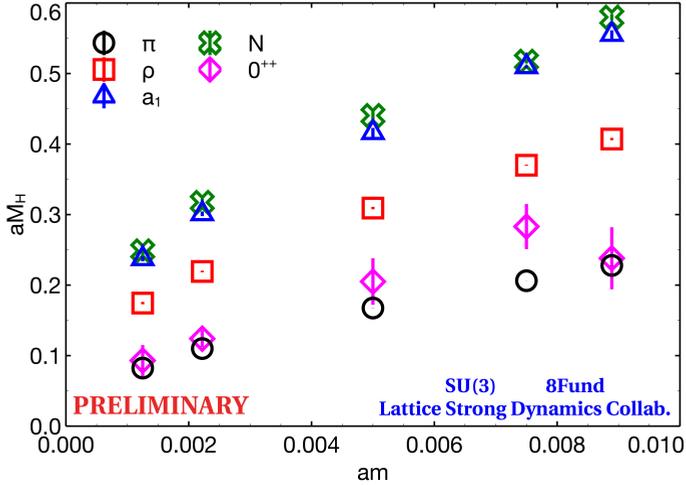}
  \caption{Hadron spectra (in lattice units) for the SU(3) theory with $N_f=8$.
  Updated data from the LSD collaboration \cite{Fleming}.}
  \label{fig:LSD1}
\end{figure}

The picture is different for truly small values of $m_f$.
Figure~\ref{fig:LSD1} shows recent data from LSD taken at very small mass values.
%%%%%%%%%%%%%%%%%%%%%%%%%%%%%%%%%%%%%%%%%%%%%%%%%%%%%%%%%%%%%%%%%%%%%%%%%%%
%%%%%%%%%%%%%%%%%%%%%%%%%%%%%%%%%%%%%%%%%%%%%%%%%%%%%%%%%%%%%%%%%%%%%%%%%%%
It looks like $m_\pi$ tends to zero as $m_f\to0$ while other masses ($\rho$, $a_1$, and the nucleon) do not.
The surprise is the scalar, which tracks the pseudoscalar pion towards zero!
It would be nice to get a scalar that is light but {\em not quite\/} massless.

\begin{figure}[hbt]
  \centering
  %\sidecaption
  \includegraphics[width=9cm,clip]{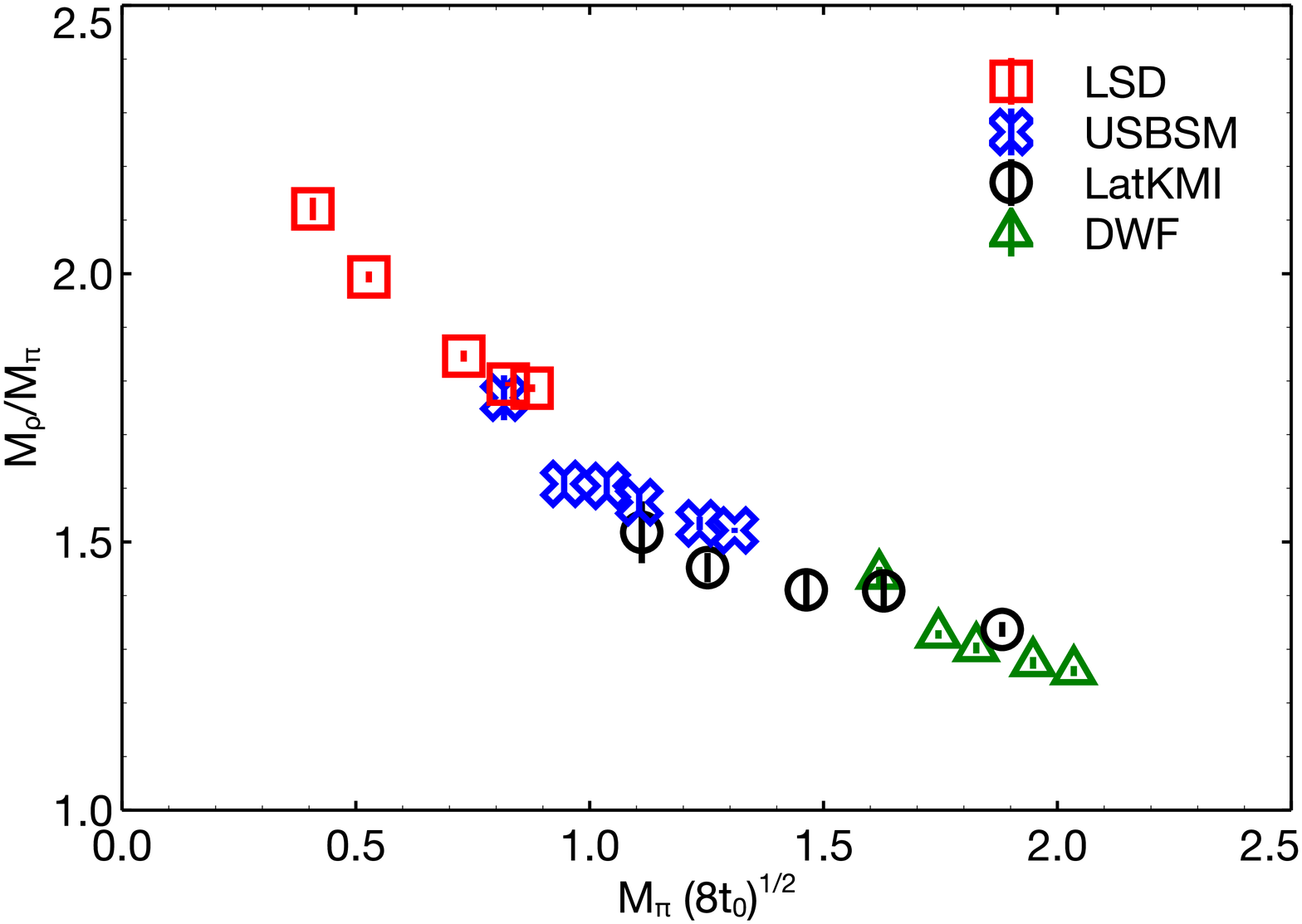}
  \caption{Edinburgh plot for the data shown in Fig.~\ref{fig:LSD1} (red squares), together with older LSD (DWF) data~\cite{Appelquist:2014zsa} and LatKMI results~\cite{Aoki:2016wnc}.}
  \label{fig:LSD2}
\end{figure}

While Fig.~\ref{fig:LSD1} is dramatic, one should note that it is a plot of lattice quantities.
In this and similar theories, because of the large number of fermions the lattice spacing can change rapidly as a function of fermion mass.
Thus one can get more reliable quantitative information from the Edinburgh plot shown in Fig.~\ref{fig:LSD2}, where masses are multiplied by the gradient flow scale $\sqrt{t_0}$ in order to make them physical.
%%%%%%%%%%%%%%%%%%%%%%%%%%%%%%%%%%%%%%%%%%%%%%%%%%%%%%%%%%%%%%%%%%%%%%%%%%%
%%%%%%%%%%%%%%%%%%%%%%%%%%%%%%%%%%%%%%%%%%%%%%%%%%%%%%%%%%%%%%%%%%%%%%%%%%%
Indeed one sees a plot that is typical for a confining theory, with $M_\rho$ staying finite as $M_\pi$ tends to zero.

%%%%%%%%%%%%%%%%%%%%%%%%%%%%%%%%%%%%%%%%%%%%%%%%%%%%%%%%%%%%%%%%%%%%%%%%%%%
\section{Forcing a theory to walk: $N_f=4\ell+8h$}
%%%%%%%%%%%%%%%%%%%%%%%%%%%%%%%%%%%%%%%%%%%%%%%%%%%%%%%%%%%%%%%%%%%%%%%%%%%
A theory with 8 flavors might make you uneasy: When chiral symmetry breaks spontaneously, it drops 63 Goldstone bosons on us.
In technicolor, three of these get eaten by the electroweak vector bosons.
Of the remainder, 36 are neutral under the electroweak gauge group---so they have to go somewhere.
Moreover, when 8 flavors run around vacuum polarization diagrams they play havoc with the Peskin--Takeuchi parameters $S,T,U$, already constrained by ``low-energy'' precision experiments.

An interesting and simple mechanism for dealing with these problems is to give the unwanted fermion species an explicit mass in a way that they can still be used to induce near-conformality and walking.
An example of this sort of thing, still amenable to staggered-fermion simulations, is an SU(3) gauge theory with 4 light ($\ell$) and 8 heavy ($h$) flavors \cite{Brower:2015owo,Hasenfratz:2016gut,Rebbi}.
If the theory is near-conformal at some scale $\Lambda$, one lifts the 8 flavors with a mass $m_h<\Lambda$, leaving 4 flavors for the low-energy theory which can be kept light, even to the limit $m_\ell\to0$.
The near-conformality is supposed to result in a light scalar as above.

What's unpleasant about this is that $m_h$ has to come from somewhere, presumably from a yet higher energy scale, much like the SM fermion masses in technicolor.
But ``one thing at a time.''
Similarly, $N_\ell=4$ still leaves too many Goldstone bosons for technicolor.
Perhaps, as claimed by the authors of this mechanism, this theory can find a home as a more general Composite Higgs model---with a light Higgs, to be sure.

\begin{figure}[hbt]
  \centering
  %\sidecaption
\includegraphics[width=4.5cm,clip,origin=c]{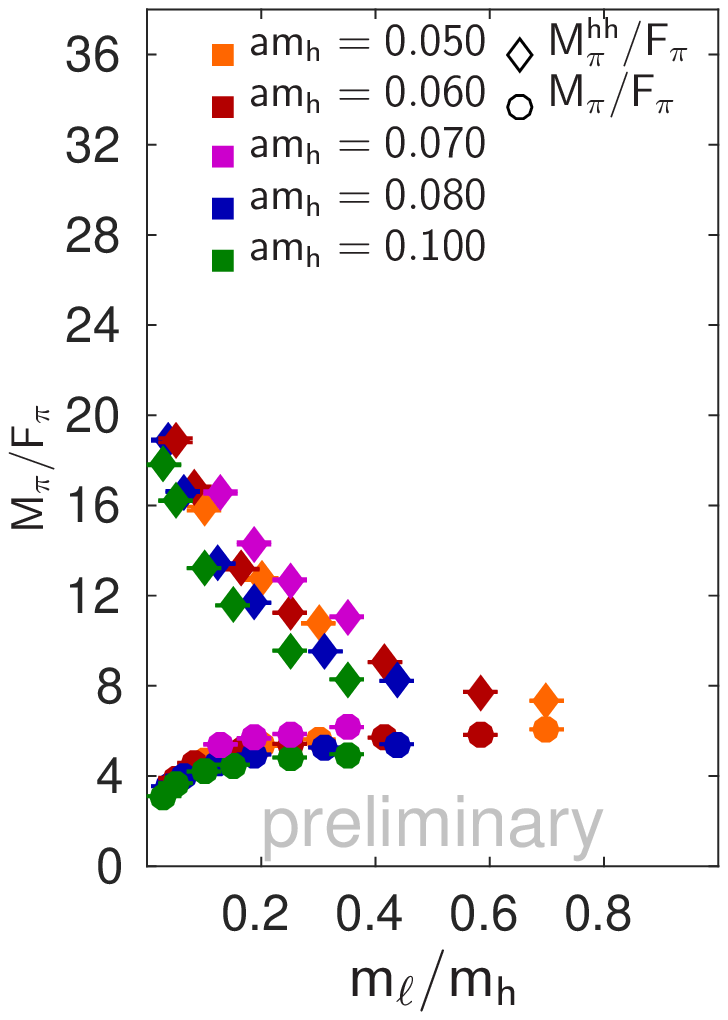}~~%
\includegraphics[width=4.5cm,clip,origin=c]{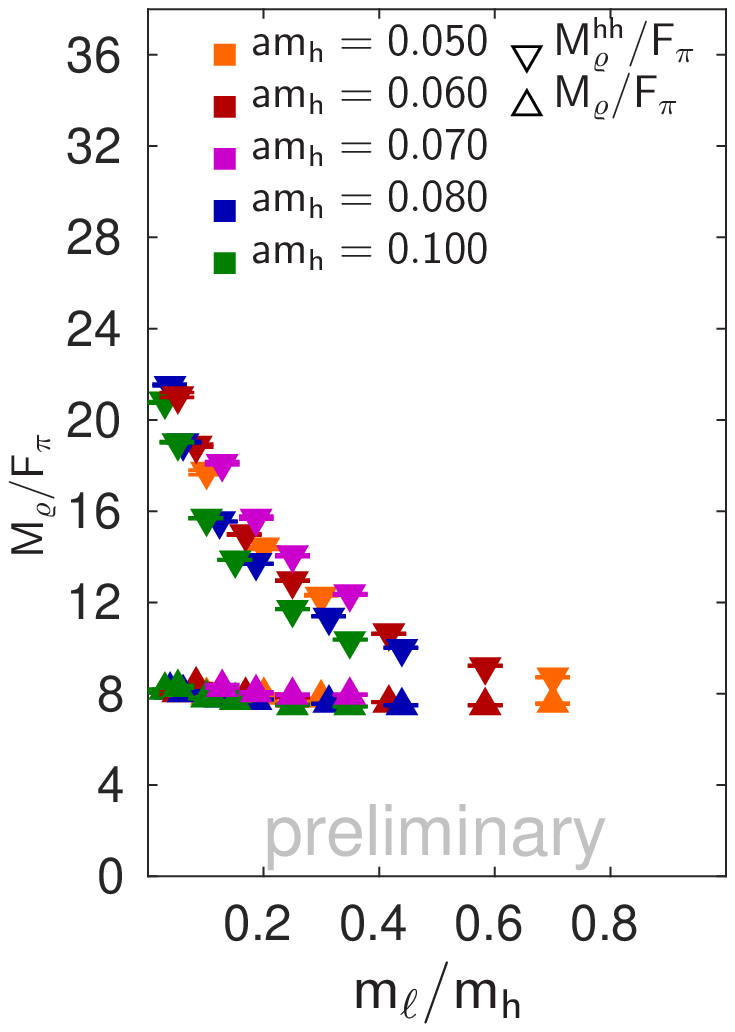}~~%
\includegraphics[width=4.5cm,clip,origin=c]{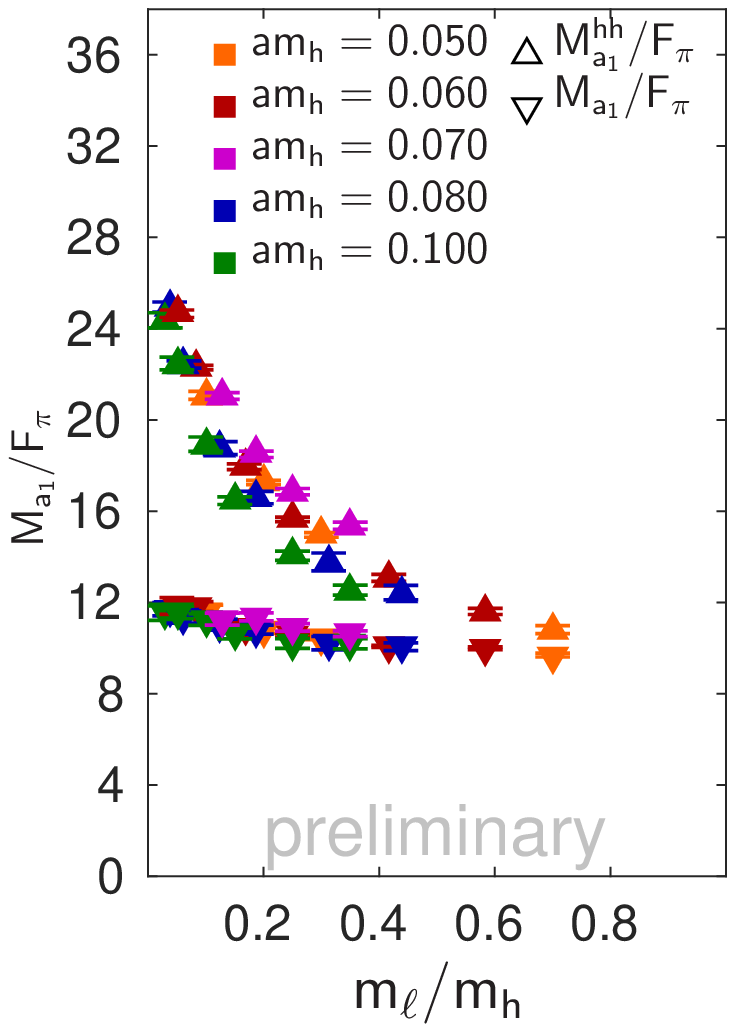}%
  \caption{Hyperscaling in the SU(3) gauge theory with 4 light and 8 heavy fermion flavors \cite{Rebbi}.}
  \label{fig:Witzel}
\end{figure}

To see that the mechanism works, one examines the masses of mesons made of $\ell\ell$ and $hh$ quark pairs as functions of $m_\ell$ and $m_h$---see Fig.~\ref{fig:Witzel}.
%%%%%%%%%%%%%%%%%%%%%%%%%%%%%%%%%%%%%%%%%%%%%%%%%%%%%%%%%%%%%%%%%%%%%%%%%%%
%%%%%%%%%%%%%%%%%%%%%%%%%%%%%%%%%%%%%%%%%%%%%%%%%%%%%%%%%%%%%%%%%%%%%%%%%%%
These show hyperscaling, inherited from the nearby IR fixed point.
For one thing, the masses are functions only of the dimensionless ratio $m_\ell/m_h$, over a wide range of $m_h$, as can be seen for the $\pi$, $\rho$, and $a_1$, respectively, in the three parts of the figure.
It is remarkable that this scaling applies even to the $hh$ mesons.
It is also a fact that the data in Fig.~\ref{fig:Witzel} were taken at more than one value of the gauge coupling $\beta$, which is indicative of the expected irrelevance of $\beta$ at the conformal fixed point.
So the entire spectrum shows that there is a fixed point where the theory walks for a range of scales.
To be sure, however, the theory tears itself away from the fixed point in the IR, as can be seen by the fact that  $\ell\ell$ pion moves towards zero mass as $m_\ell/m_h\to0$.
This is a sign of a chiral condensate.
Not shown in the figures is the fact that there is a light scalar whose mass tracks that of the pion; again, one may ask where, exactly, this scalar ends up.
%%%%%%%%%%%%%%%%%%%%%%%%%%%%%%%%%%%%%%%%%%%%%%%%%%%%%%%%%%%%%%%%%%%%%%%%%%%
\section{Is the Higgs a pseudo-dilaton?\label{sec:dilaton}}
%%%%%%%%%%%%%%%%%%%%%%%%%%%%%%%%%%%%%%%%%%%%%%%%%%%%%%%%%%%%%%%%%%%%%%%%%%%
Having found a light scalar in several theories with walking, we might hope for an explicit sign that it is indeed the walking---the approximate scale invariance---that brings it about.
This is important because the Higgs isn't just any light scalar.
It has to couple to SM particles in proportion to their masses, a feature of the Weinberg--Salam scalar field that is so far confirmed by LHC data.

The dilatation current in a gauge theory is anomalous and thus satisfies an equation of partial conservation,
\begin{equation}
\partial_\mu S^\mu=T^\mu_\mu=-\frac{\beta(g^2)}{4g^2}F^2-(1+\gamma_m)\,m\,\bar\psi\psi.
\end{equation}
Conservation of the current is broken by the fermion mass $m$ and by the beta function $\beta(g^2(\Lambda))$, which we assume to be parametrically small.

If we were talking about a partially conserved chiral symmetry, we would have a Gell-Mann--Oakes--Renner relation,
\begin{equation}
m_\pi^2f_\pi^2=m_q\langle\bar\psi\psi\rangle,
\end{equation}
giving the mass of the pseudo-Goldstone boson in an expansion around the symmetric point
$m_q=0$.
It is not obvious that there is an analogue of this expansion for the dilaton.
After all, if $\beta=m=0$ then the theory is conformal, and there are {\em no\/} Goldstone bosons, contrary to the chiral example where there is an exactly massless pion in this limit.
It is therefore something of a leap to suppose that an effective action can be written down for the spontaneous breaking of chiral and dilatation symmetries, and that this effective action allows a systematic expansion around a symmetric point.
Golterman and Shamir did exactly this \cite{Golterman:2016lsd,Golterman:2016hlz}, through an expansion in a small parameter $\delta$ that is supposed to characterize the usual chiral expansion,
\begin{equation}
p^2/\Lambda^2\ \sim\ m/\Lambda\ \sim\ \delta,
\end{equation}
as well as the distance below the sill of the conformal window (where dilatation symmetry is exact),%
\footnote{In order to make \Eq{delta} work, they formally work in a large-$N_c$ regime by assuming as well that $1/N_c\sim\delta$.
The power counting is constructed for fermions in the fundamental representation of the gauge group.}
\begin{equation}
|N_f-N_f^*|/N_c \ \sim \ \delta.
\label{delta}
\end{equation}
The action for the dilaton field $\tau$ takes the form
\begin{equation}
{\cal L}_\tau =
  \frac{f_\tau^2}{2}\, V_\tau(\tau)\, e^{2\tau} (\partial_\mu \tau)^2  
  +f_\tau^2 B_\tau \, V_d(\tau)\, e^{4\tau} +\textit{coupling to}\ \pi.
\label{dilaton}
\end{equation}
The rest of the expression is the usual pion effective action, rendered scale-invariant by the necessary factors of $e^\tau$ and then non-invariant by further functions similar to $V_\tau,V_d$.
The potentials $V$ can be expanded systematically in powers of the breaking parameter $\delta$, e.g.,
\begin{equation}
V_d=\sum c_n\tau^n, \quad c_n=O(\delta^n).
\end{equation}
As is done for chiral lagrangians, one fixes the constants in this action (including the dilaton mass $m_\tau$) from correlators in the gauge theory.
(See also \cite{Kasai:2016ifi,Golterman:2016cdd}.)

Fits to a dilaton effective action have been carried out for the 8-flavor theory by LatKMI \cite{Aoki:2016wnc} as well as members of LSD \cite{Appelquist:2017wcg,Gasbarro}.
These fits have (mostly) used a lagrangian with {\em ad hoc\/} dilatation-breaking terms, not the full systematic expansion of \Eq{dilaton}, but they are quite successful and show the necessity of including the dilaton as a low-energy excitation.

%%%%%%%%%%%%%%%%%%%%%%%%%%%%%%%%%%%%%%%%%%%%%%%%%%%%%%%%%%%%%%%%%%%%%%%%%%%
\begin{figure}[hbt]
  \centering
  %\sidecaption
  \includegraphics[width=9cm,clip]{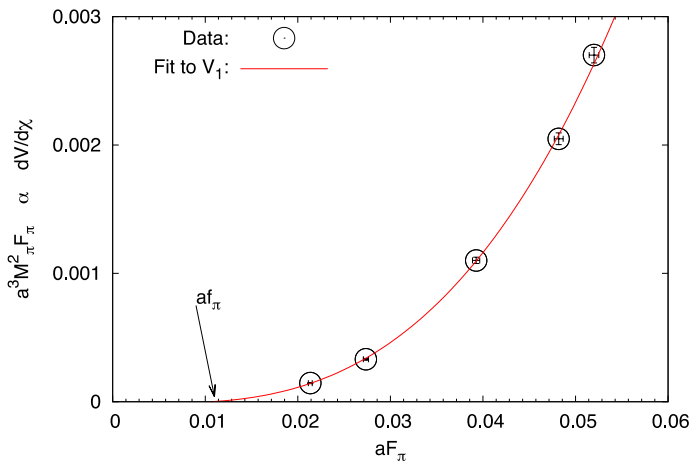}
  \caption{Fit to LSD data employing a dilaton effective action.}
  \label{fig:dilaton}
\end{figure}
%%%%%%%%%%%%%%%%%%%%%%%%%%%%%%%%%%%%%%%%%%%%%%%%%%%%%%%%%%%%%%%%%%%%%%%%%%%
Figure~\ref{fig:dilaton} shows a fit to the LSD data, plotting $m_\pi^2F_\pi$ against $F_\pi$ (the reason for this plot can be found in \cite{Appelquist:2017wcg}).
Apart from the good quality of the fit, what I want to point out first is that 
$F_\pi$ (the abscissa) varies by a factor of 2.4 across the data, which would normally spell trouble for a fit to a pure chiral lagrangian.
Evidently, if the dilaton is practically degenerate with the pion as in Fig.~\ref{fig:LSD1}, you had better include it in the low-energy effective theory.

Second, the extrapolation to the chiral limit along a steep curve then lowers $F_\pi$ by a further factor of two to its limit $f_\pi$.
This may be a mechanism for obtaining a Higgs vev, proportional to $f_\pi$, in the physical range despite a technicolor scale of $\Lambda\sim5$~TeV.

%%%%%%%%%%%%%%%%%%%%%%%%%%%%%%%%%%%%%%%%%%%%%%%%%%%%%%%%%%%%%%%%%%%%%%%%%%%
\section{Other technicolor models\label{sec:other}}
%%%%%%%%%%%%%%%%%%%%%%%%%%%%%%%%%%%%%%%%%%%%%%%%%%%%%%%%%%%%%%%%%%%%%%%%%%%

I have presented some of the dynamical issues in the application of technicolor to the quest for a light Higgs boson.
These have also been explored in other gauge theories.
I will just list the work that has been done recently.
\begin{itemize}
\item The SU(2) gauge theory with 8 flavors in the fundamental representation appears to be in the conformal window.
This is a result of calculating the running coupling defined through the gradient flow, in the lattice theory with smeared Wilson fermions \cite{Leino:2017lpc}.
\item The SU(2) gauge theory with two flavors in the {\em adjoint\/} representation is conformal, according to the results of many studies; a recent study of its spectrum with smeared Wilson fermions confirms this \cite{Bergner:2016hip}.
If one adds a flavor-breaking four-fermi coupling, the theory moves into a phase where the remnant chiral symmetry is spontaneously broken \cite{Rantaharju:2017eej};
knowledge of the spectrum here is incomplete.
\item The SU(3) gauge theory with 2 flavors in the sextet representation is an old friend.
It has a long history that I won't relate, but I'll note that extensive calculations of the spectrum \cite{Fodor:2012ty,Wong} and of the gradient-flow running coupling \cite{Fodor:2015zna,Holland} are cited as evidence that this is a walking theory with a light Higgs (see also \cite{Fodor:2016wal}). 
A dilaton effective action has also found application here \cite{Kuti}.
These calculations have all used rooted staggered fermions.
A recent calculation of the spectrum with (unsmeared) Wilson fermions, on the other hand, places the theory in the conformal window \cite{Hansen:2017ejh,Pica}.
\end{itemize}
Apart from the questions that I have dealt with above, other quantities of phenomenological interest are calculable with lattice methods.
Among these are the anomalous dimension $\gamma_m$ of the mass operator, needed for extended technicolor scenarios of SM fermion masses; and how well these theories can satisfy precision tests, quantified mainly via the $S$ parameter to which I alluded before.

%%%%%%%%%%%%%%%%%%%%%%%%%%%%%%%%%%%%%%%%%%%%%%%%%%%%%%%%%%%%%%%%%%%%%%%%%%%
\section{The Higgs as a composite pseudo-Goldstone boson \label{sec:CH}}
%%%%%%%%%%%%%%%%%%%%%%%%%%%%%%%%%%%%%%%%%%%%%%%%%%%%%%%%%%%%%%%%%%%%%%%%%%%
So far we've been discussing technicolor, which uses a strong coupling theory with scale $\Lambda\simeq 5$~TeV to break the electroweak gauge group spontaneously, with a miraculous mechanism (like broken dilatation symmetry) to protect the Higgs mass and the weak scale, putting them in the physical range $\sim\Lambda/20$.
An alternative proposal is to generate the Higgs multiplet as massless, composite Goldstone fields, to be given a vev $v$ and a mass by coupling to the Standard Model \cite{Georgi:1984af,Dugan:1984hq}.
Thus we start with a {\em hypercolor\/} theory with scale $f\gg v$ that breaks its global symmetry spontaneously in the usual way, leaving unbroken a subgroup that includes the SU(2)$_L\,\times\,$U(1) of the SM.
The Higgs multiplet $h$ is hence included among the Goldstone bosons, so $m_H=0$ and in fact there is no Higgs potential at all.
Now we couple the theory to the gauge bosons and fermions of the SM.
One-loop diagrams then generate a Higgs potential, generically of the form \cite{Golterman:2015zwa,Golterman:2017vdj}
\begin{equation}
V_{\rm eff}(h)=\alpha\cos(2h/f)-\beta\sin^2(2h/f).
\label{Higgsaction}
\end{equation}
The coefficient $\alpha$ has a piece due to a gauge boson loop and a piece due to a top-quark loop,
\begin{equation}
\alpha=-\frac12(3g^2+g^{\prime2})C_{LR}+\alpha_{\textrm{top}},
\end{equation}
while the other coefficient $\beta$ is due entirely to a top loop.
The coefficient $C_{LR}$ is an integral of a current--current correlator in the hypercolor theory, which is fairly straightforward to calculate \cite{DeGrand:2016htl}.

If $\alpha+2\beta>0$ then $h=0$ is unstable.
This breaks the electroweak symmetry spontaneously, giving a Higgs vev $v=\sqrt2\langle h\rangle$ according to
\begin{equation}
\cos(\sqrt{2}v/f) = -\alpha/(2\beta).
\end{equation}
(If we demand $v/f\ll1$ then we expand to obtain $(v/f)^2\approx 1+\alpha/(2\beta)$.
This demands a miraculous tuning of $\alpha/\beta$.
Maybe there is a mechanism to bring this about.
Maybe not.)

%%%%%%%%%%%%%%%%%%%%%%%%%%%%%%%%%%%%%%%%%%%%%%%%%%%%%%%%%%%%%%%%%%%%%%%%%%%
\section{And a partially composite top quark \label{sec:PCT}}
%%%%%%%%%%%%%%%%%%%%%%%%%%%%%%%%%%%%%%%%%%%%%%%%%%%%%%%%%%%%%%%%%%%%%%%%%%%
There is a long list of models that embody the pseudo-Goldstone mechanism for the Higgs.
Phenomenologists have generally defined them by the low-energy Lagrangian that describes the symmetry-breaking scheme that creates the Goldstone Higgs field, and by its subsequent coupling to the Standard Model; this is typically a non-renormalizable, nonlinear sigma model.%
\footnote{Since this literature may be less familiar than that of technicolor, I'll cite some reviews \cite{Contino:2010rs,Bellazzini:2014yua,Panico:2015jxa}.}
Ferretti and Karateev \cite{Ferretti:2013kya} winnowed down this list by imposing two reasonable constraints.
One is that the proposed symmetry-breaking should be consistent with the low-energy physics of an asymptotically free gauge theory---the {\em UV completion\/} of the sigma model.
This requirement is obvious to lattice gauge theorists!

The other requirement is a violation of my {\em one thing at a time\/} principle:
While we're solving the problem of a natural Higgs boson, let's offer an explanation of the 172~GeV mass of the top quark.
The top is similar to the Higgs in that it is much heavier than the other quarks but much lighter than any practical compositeness scale $\Lambda\!\gtrsim\!5$~TeV.
A possible mechanism for the top mass is Kaplan's {\em partial compositeness\/} \cite{Kaplan:1991dc}, in which a massless, fundamental $t$ quark mixes with a composite $T$ particle that emerges from the hypercolor scale and hence lives at the $\Lambda$ scale.
Generically, this mixing can result in a physical $t$ quark in the right range, well below $\Lambda$.%
\footnote{In the SM with a fundamental Higgs field, the top quark requires a strong Yukawa coupling to produce its mass.
Technicolor is largely silent on the subject, because if such a large mass is generated through extended technicolor then there will be large flavor-changing neutral current vertices as well.}

An attractive model that emerges from these considerations is an SU(4) gauge theory with fermions in two representations---a {\em multirep\/} theory \cite{Ferretti:2014qta}.
First, there are sextet fermions.
The sextet is the two-index antisymmetric representation of SU(4), and it is a real representation.
The choice of 5 Majorana flavors $Q_a$ in the hypercolor sextet gives the theory an SU(5) chiral symmetry, that will break spontaneously to SO(5) upon the appearance of a bilinear condensate.
The SU(5)/SO(5) coset turns out to offer a good embedding of the Higgs multiplet.

To produce a top partner, one adds 3 flavors of Dirac fermions $q_j$ in the fundamental representation of hypercolor.
Then a hyper-colorless baryon state $T$ can be formed as a bound state $Qqq$---a {\em chimera\/} of the two fermion species.
If one gauges the flavor index of the 3 Dirac fermions, the chimera baryon becomes a color triplet.
Its other quantum numbers are as needed to mix with a fundamental $t$ quark.

%%%%%%%%%%%%%%%%%%%%%%%%%%%%%%%%%%%%%%%%%%%%%%%%%%%%%%%%%%%%%%%%%%%%%%%%%%%
\section{Multirep \label{sec:multirep}}
%%%%%%%%%%%%%%%%%%%%%%%%%%%%%%%%%%%%%%%%%%%%%%%%%%%%%%%%%%%%%%%%%%%%%%%%%%%
Multirep theories open a new dimension in the study of gauge dynamics and, in particular, in lattice gauge theory.
Apart from the influence of each fermion species on the gauge field and vice versa, phase transitions and symmetry breaking in each species can affect the others dramatically.
Of course, QCD already contains light quarks, strange quarks, and heavy quarks, and the influence of each species on the others is an old and continuing object of QCD calculations.
The difference is that QCD's quarks are all equivalent, in that a tuning of the masses can change one into another.
Fermions in inequivalent representations, on the other hand, enter the dynamics with different strengths even if the masses are made degenerate.

If all the fermions are made massless, the chiral symmetries of the species remain distinct.
One symmetry could break spontaneously while others do not.
This is a generalization of the old issue of scale separation, which was originally seen as a possible separation of a chiral scale from the confinement scale of the gauge theory.
It is possible that inequivalent representations, simultaneously coupled to the gauge field, define independent scales.
This might find expression in the finite-temperature physics of the theory, in the form of distinct phase transitions for each fermion species as well as for the confinement physics of the gauge field.
Alternatively, one phase transition might trigger all the others to occur at the same scale.

\begin{figure}[hbt]
  \centering
  %\sidecaption
  \includegraphics[width=11cm,clip]{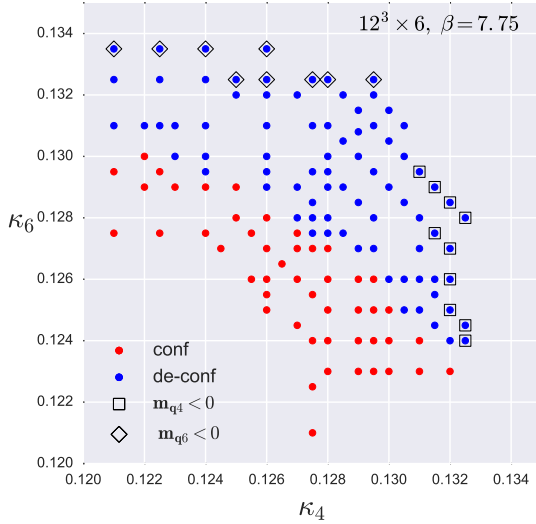}
  \caption{Phase diagram of the multirep SU(4) gauge theory at fixed lattice coupling $\beta$ \cite{Venkitesh}.
  The axes are $\kappa_4$ and $\kappa_6$, the hopping parameters of fermions in the fundamental and sextet representations, respectively.}
  \label{fig:multirep}
\end{figure}

To see that multirep is a whole new game in lattice gauge theory, one need only glance at Fig.~\ref{fig:multirep} from the work of the TACO collaboration%
\footnote{$\ldots$ of which I am a member. I'll keep the discussion in the third person.}
 \cite{Will,Venkitesh,Dan}.
%%%%%%%%%%%%%%%%%%%%%%%%%%%%%%%%%%%%%%%%%%%%%%%%%%%%%%%%%%%%%%%%%%%%%%%%%%%
%%%%%%%%%%%%%%%%%%%%%%%%%%%%%%%%%%%%%%%%%%%%%%%%%%%%%%%%%%%%%%%%%%%%%%%%%%%
The multirep theory here is close to that described above---likewise an SU(4) gauge theory with fermions in the fundamental and sextet representations of the gauge group.
To simplify the simulations, TACO chose to include 2 flavors of fundamental fermions (instead of 3) and 2 Dirac flavors of sextet fermions (equivalent to 4 Majorana flavors, instead of 5).
The bare-coupling space is three-dimensional: the gauge coupling $\beta$ and two hopping parameters $\kappa_4$ and $\kappa_6$ for the two inequivalent fermion species.
Figure~\ref{fig:multirep} is a phase diagram in the $(\kappa_4,\kappa_6)$ plane at fixed $\beta$. The  $\kappa_c$ of each species is a function of the $\kappa$ of the other species, as can be seen from the diamonds and squares.
The boundary between red and blue points is the finite-temperature confinement transition on this $N_t=6$ lattice; it appears that there is only one transition, so there is no smoking gun for scale separation.
Let me point out that while this plot represents a plane at fixed $\beta$, it is far from representing a fixed lattice spacing, since TACO finds that the lattice scale $t_0$ changes rapidly with variation of $\kappa_4$ and $\kappa_6$.
Working in a three-dimensional coupling space can be tedious, but as I said, it's a new game.%
\footnote{Preliminary work on the chimera baryon was presented by W. I. Jay \cite{DeGrand:2016mxr} last year.}

An alternative model under study is based on a multirep Sp(4) gauge theory, originally proposed in  \cite{Barnard:2013zea}.
Several talks at this conference have presented preliminary results regarding the composite Higgs aspect of this model \cite{Lucini,Vadacchino,Bennett}, but so far only one species of fermion (in the fundamental representation) has been included and so there is no multirep physics as yet---and no baryon for partial compositeness, either.
%%%%%%%%%%%%%%%%%%%%%%%%%%%%%%%%%%%%%%%%%%%%%%%%%%%%%%%%%%%%%%%%%%%%%%%%%%%
\section{Phenomenology? \label{sec:pheno}}
\begin{figure}[hbt]
  \centering
  %\sidecaption
  \includegraphics[width=8cm,clip]{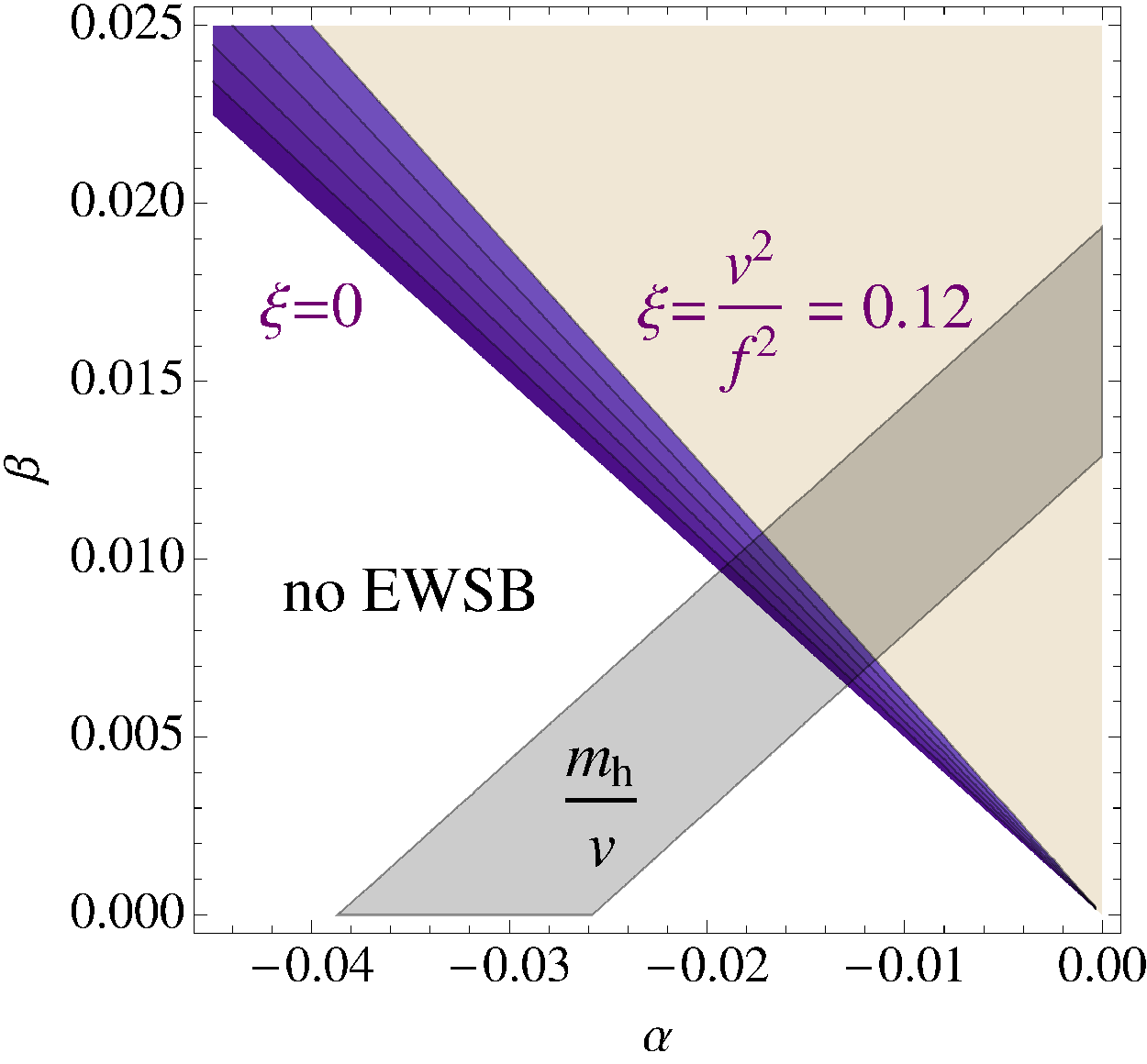}
  \caption{Constraints on the coefficients of the effective Lagrangian (\ref{Higgsaction}) from existence and mass of the Higgs.}
  \label{fig:Higgsconstraints}
\end{figure}

%%%%%%%%%%%%%%%%%%%%%%%%%%%%%%%%%%%%%%%%%%%%%%%%%%%%%%%%%%%%%%%%%%%%%%%%%%%
Let me close by mentioning a recent attempt \cite{DelDebbio:2017ini} to constrain the SU(4) composite Higgs model with present and future LHC data.
One result is that the coefficients $\alpha$ and $\beta$ in \Eq{Higgsaction} are bounded by straightforward statements about the Higgs boson, as seen in Fig.~\ref{fig:Higgsconstraints}.
%%%%%%%%%%%%%%%%%%%%%%%%%%%%%%%%%%%%%%%%%%%%%%%%%%%%%%%%%%%%%%%%%%%%%%%%%%%
%%%%%%%%%%%%%%%%%%%%%%%%%%%%%%%%%%%%%%%%%%%%%%%%%%%%%%%%%%%%%%%%%%%%%%%%%%%
I have already mentioned the requirement $\alpha+2\beta>0$ that comes of the requirement of a nonzero Higgs vev $v$.
This eliminates the white part of the figure.
The purple area consists of rays fixed by various ratios of $\xi\equiv v^2/f^2$, where $f$ is the chiral parameter (the pion decay costant) of the strongly coupled hypercolor theory, related to its scale $\Lambda$.
Evidently, the higher the strong scale, the closer one moves to $\xi=0$.
The other constraint plotted comes from the measured Higgs mass, determined by \Eq{Higgsaction} as
\begin{equation}
m^2_h/v^2=8(2\beta-\alpha).
\end{equation}
This is the gray stripe across the figure.

The nagging problem with such constraints is that the coefficients $\alpha$ and $\beta$ contain top quark contributions.
These depend on the top quark's Yukawa coupling $y_t$, which in turn comes from the calculable coupling of the top quark partner---the (composite) chimera baryon of the hypercolor model---to the composite Higgs  \cite{Golterman:2015zwa,Golterman:2017vdj}.
Unfortunately, relating the two involves knowing the mixing parameters of the fundamental top quark $t$ with the composite partner $T$.
This mixing is expressed by an effective Lagrangian, generically of the form
\begin{equation}
 {\cal L}_{\rm EHC} =  {\lambda_1} (\bar T_L t_R + \bar t_R T_L)
 + {\lambda_2} (\bar T_R t_L + \bar t_L T_R).
 \end{equation}
The subscript EHC indicates that this mixing Lagrangian has to emerge from an ``extended'' hypercolor theory, defined at some unknown scale in order to produce effective four-fermi interactions between $t$ and the hyperquarks.
Given the coefficients $\lambda_1,\lambda_2$, a lattice calculation similar to the calculation of $C_{LR}$ will give the Yukawa coupling $y_t$ and hence the coefficients $\alpha,\beta$.
At the hypercolor level, however, the mixing parameters come from a black box.
{\em One thing at a time.}

The model has a rich particle content.
There are exotic Goldstone bosons, among them some with QCD interactions.
The authors of Ref.~\cite{DelDebbio:2017ini} have plotted what future limits can be set by the LHC on the production of these particles.
There is a minus, though:
The interaction of some Goldstone bosons with the Higgs field affects the $\rho$ parameter
\cite{Golterman:2017vdj}.
This might spell the doom of this theory as a practical model of the Higgs and top quark.

%%%%%%%%%%%%%%%%%%%%%%%%%%%%%%%%%%%%%%%%%%%%%%%%%%%%%%%%%%%%%%%%%%%%%%%%%%%
\section{Summary \label{sec:summary}}
%%%%%%%%%%%%%%%%%%%%%%%%%%%%%%%%%%%%%%%%%%%%%%%%%%%%%%%%%%%%%%%%%%%%%%%%%%%
Looking back on what I've covered, I can reduce the material to a few nutshells.
\begin{enumerate}
\item{Technicolor}
\begin{itemize}
\item
The effort to nail down the sill of the conformal window continues, for several gauge groups and fermion representations (one representation at a time).
I discussed the case of the SU(3) gauge theory with fundamental fermions: $N_f=12$ might or might not be above the sill.
\item
If you know you're below the sill, it makes sense to look for walking as a mechanism for a light scalar that might be the Higgs boson.
An example is SU(3) with $N_f=8$.
Or you can try to make a theory walk by starting above the sill at high energy and dropping below it at low energy, as in the $N_f=4\ell+8h$ model.
Here you need an excuse for the heavy quarks' mass term.
\item
Has your walking theory really produced a light, dilatonic Higgs and a low scale for the Higgs vev?
Check this by matching your results to a dilatonic effective action.
An ordinary chiral model is {\em not\/} good enough.%
\footnote{ And please show me that the Higgs mass is nonzero in the chiral limit.}
\end{itemize}
\item{Composite Higgs and partially composite top quark}
\begin{itemize}
\item Multirep models are a whole new area for lattice simulations.
\item There are many opportunities for lattice calculations of low-energy constants.
Unfortunately, they will always depend on unknown mixing parameters that come from yet higher energies.
Still, calculations might yield some surprising systematics.
\item Phenomenological constraints on these theories might be premature, since the models aren't perfect.
\end{itemize}
So my conclusion is that there's lots of interesting work to do, even if the environment is somewhat obscure and even threatening.
\end{enumerate}

%%%%%%%%%%%%%%%%%%%%%%%%%%%%%%%%%%%%%%%%%%%%%%%%%%%%%%%%%%%%%%%%%%%%%%%%%%%
\section*{Acknowledgements}
%%%%%%%%%%%%%%%%%%%%%%%%%%%%%%%%%%%%%%%%%%%%%%%%%%%%%%%%%%%%%%%%%%%%%%%%%%%
I thank David Schaich and Oliver Witzel for sending me unpublished data, as well as my collaborator Venkitesh Ayyar for Fig.~\ref{fig:multirep}.
I am grateful for Yigal Shamir's assistance in preparing this lecture.

%%%%%%%%%%%%%%%%%%%%%%%%%%%%%%%%%%%%%%%%%%%%%%%%%%%%%%%%%%%%%%%%%%%%%%%%%%%
%\bibliography{lattice2017}

\begin{thebibliography}{99}
%%%%%%%%%%%%%%%%%%%%%%%%%%%%%%%%%%%%%%%%%%%%%%%%%%%%%%%%%%%%%%%%%%%%%%%%%%%

\bibitem{Oyanguren}
A. Oyanguren [LHCb Collaboration], plenary lecture at this conference:
  %``B decay anomalies at LHCb,''
  EPJ Web Conf.\  {\bf 175}, 01004 (2018).
  %%CITATION = doi:10.1051/epjconf/201817501004;%%
\bibitem{Becirevic}
D. Be\v cirevi\'{c}, plenary lecture at this conference.


\bibitem{DeGrand:2015zxa} 
  T.~DeGrand,
  %``Lattice tests of beyond Standard Model dynamics,''
  Rev.\ Mod.\ Phys.\  {\bf 88}, 015001 (2016)
  [arXiv:1510.05018 [hep-ph]].
  %%CITATION = doi:10.1103/RevModPhys.88.015001;%%
  
\bibitem{Nogradi:2016qek} 
  D.~Nogradi and A.~Patella,
  %``Strong dynamics, composite Higgs and the conformal window,''
  Int.\ J.\ Mod.\ Phys.\ A {\bf 31}, 1643003 (2016)
  [arXiv:1607.07638 [hep-lat]].
  %%CITATION = doi:10.1142/S0217751X1643003X;%%

\bibitem{Pica:2017gcb} 
  C.~Pica,
  %``Beyond the Standard Model: Charting Fundamental Interactions via Lattice Simulations,''
  PoS LATTICE {\bf 2016}, 015 (2016)
  [arXiv:1701.07782 [hep-lat]].
  %%CITATION = ARXIV:1701.07782;%%

\bibitem{Lin:2015zpa} 
  C.-J.~D.~Lin, K.~Ogawa, and A.~Ramos,
  %``The Yang-Mills gradient flow and SU(3) gauge theory with 12 massless fundamental fermions in a colour-twisted box,''
  JHEP {\bf 1512}, 103 (2015)
  [arXiv:1510.05755 [hep-lat]].
  %%CITATION = doi:10.1007/JHEP12(2015)103;%%

\bibitem{Fodor:2016zil} 
  Z.~Fodor, K.~Holland, J.~Kuti, S.~Mondal, D.~Nogradi, and C.~H.~Wong,
  %``Fate of the conformal fixed point with twelve massless fermions and SU(3) gauge group,''
  Phys.\ Rev.\ D {\bf 94}, 091501 (2016)
  [arXiv:1607.06121 [hep-lat]].
  %%CITATION = doi:10.1103/PhysRevD.94.091501;%%
  
\bibitem{Hasenfratz:2016dou} 
  A.~Hasenfratz and D.~Schaich,
  JHEP {\bf 1802}, 132 (2018)
  [arXiv:1610.10004 [hep-lat]].
  %%CITATION = doi:10.1007/JHEP02(2018)132;%%
  
\bibitem{Fodor}
Z. Fodor, poster at this conference: Z.~Fodor, K.~Holland, J.~Kuti, D.~Nogradi and C.~H.~Wong,
  %``The twelve-flavor $\boldsymbol{\beta}$-function and dilaton tests of the sextet scalar,''
  EPJ Web Conf.\  {\bf 175}, 08015 (2018)
  [arXiv:1712.08594 [hep-lat]].
  %%CITATION = doi:10.1051/epjconf/201817508015;%%

\bibitem{Anna}
A. Hasenfratz, lecture at this conference: A.~Hasenfratz, C.~Rebbi, and O.~Witzel,
  %``Testing Fermion Universality at a Conformal Fixed Point,''
  EPJ Web Conf.\  {\bf 175}, 03006 (2018)
  [arXiv:1708.03385 [hep-lat]].
  %%CITATION = doi:10.1051/epjconf/201817503006;%%

\bibitem{DeGrand:2010na} 
  T.~DeGrand, Y.~Shamir, and B.~Svetitsky,
  %``Running coupling and mass anomalous dimension of SU(3) gauge theory with two flavors of symmetric-representation fermions,''
  Phys.\ Rev.\ D {\bf 82}, 054503 (2010)
  [arXiv:1006.0707 [hep-lat]].
  %%CITATION = doi:10.1103/PhysRevD.82.054503;%%

\bibitem{Chiu:2016uui} 
  T.~W.~Chiu,
  %``The $\beta$-function of $SU(3)$ gauge theory with $ N_f = 10 $ massless fermions in the fundamental representation,''
  arXiv:1603.08854 [hep-lat].
  %%CITATION = ARXIV:1603.08854;%%  
  
\bibitem{Chiu:2017kza} 
  T.~W.~Chiu,
  %``Discrete $\beta$-function of the $SU(3)$ gauge theory with 10 massless domain-wall fermions,''
  PoS LATTICE {\bf 2016}, 228 (2017).
  %%CITATION = POSCI,LATTICE2016,228;%%
  
\bibitem{Aoki:2016wnc} 
  Y.~Aoki {\it et al.} [LatKMI Collaboration],
  %``Light flavor-singlet scalars and walking signals in $N_f=8$ QCD on the lattice,''
  Phys.\ Rev.\ D {\bf 96}, 014508 (2017)
  [arXiv:1610.07011 [hep-lat]].
  %%CITATION = doi:10.1103/PhysRevD.96.014508;%%
  
\bibitem{Rinaldi}
E. Rinaldi, lecture at this conference:
  Y.~Aoki {\it et al.},
  %``Flavor-singlet spectrum in multi-flavor QCD,''
  EPJ Web Conf.\  {\bf 175}, 08023 (2018)
  [arXiv:1710.06549 [hep-lat]].
  %%CITATION = doi:10.1051/epjconf/201817508023;%%

\bibitem{Appelquist:2014zsa} 
  T.~Appelquist {\it et al.} [LSD Collaboration],
  %``Lattice simulations with eight flavors of domain wall fermions in SU(3) gauge theory,''
  Phys.\ Rev.\ D {\bf 90}, 114502 (2014)
  [arXiv:1405.4752 [hep-lat]].
  %%CITATION = doi:10.1103/PhysRevD.90.114502;%%
  
\bibitem{Appelquist:2016viq} 
  T.~Appelquist {\it et al.},
  %``Strongly interacting dynamics and the search for new physics at the LHC,''
  Phys.\ Rev.\ D {\bf 93}, 114514 (2016)
  [arXiv:1601.04027 [hep-lat]].
  %%CITATION = doi:10.1103/PhysRevD.93.114514;%%

\bibitem{Fleming}
G. Fleming, lecture at this conference.

\bibitem{Brower:2015owo} 
  R.~C.~Brower, A.~Hasenfratz, C.~Rebbi, E.~Weinberg, and O.~Witzel,
  %``Composite Higgs model at a conformal fixed point,''
  Phys.\ Rev.\ D {\bf 93}, 075028 (2016)
  [arXiv:1512.02576 [hep-ph]].
  %%CITATION = doi:10.1103/PhysRevD.93.075028;%%

\bibitem{Hasenfratz:2016gut} 
  A.~Hasenfratz, C.~Rebbi, and O.~Witzel,
  %``Large scale separation and resonances within LHC range from a prototype BSM model,''
  Phys.\ Lett.\ B {\bf 773}, 86 (2017)
  [arXiv:1609.01401 [hep-ph]].
  %%CITATION = doi:10.1016/j.physletb.2017.07.058;%%
  
\bibitem{Rebbi}
C. Rebbi, lecture at this conference:   A.~Hasenfratz, C.~Rebbi and O.~Witzel,
  %``Investigating BSM Models with Large Scale Separation,''
  EPJ Web Conf.\  {\bf 175}, 08007 (2018)
  [arXiv:1710.08970 [hep-lat]].
  %%CITATION = doi:10.1051/epjconf/201817508007;%%
  
\bibitem{Golterman:2016lsd} 
  M.~Golterman and Y.~Shamir,
  %``Low-energy effective action for pions and a dilatonic meson,''
  Phys.\ Rev.\ D {\bf 94}, 054502 (2016)
  [arXiv:1603.04575 [hep-ph]].
  %%CITATION = doi:10.1103/PhysRevD.94.054502;%%
  
\bibitem{Golterman:2016hlz} 
  M.~Golterman and Y.~Shamir,
  %``Effective field theory for pions and a dilatonic mesonEffective action for pions and a dilatonic meson,''
  PoS LATTICE {\bf 2016}, 205 (2016)
  [arXiv:1610.01752 [hep-ph]].
  %%CITATION = ARXIV:1610.01752;%%
  
\bibitem{Kasai:2016ifi} 
  A.~Kasai, K.~i.~Okumura, and H.~Suzuki,
  %``A dilaton-pion mass relation,''
  arXiv:1609.02264 [hep-lat].
  %%CITATION = ARXIV:1609.02264;%%
  
\bibitem{Golterman:2016cdd} 
  M.~Golterman and Y.~Shamir,
  %``Effective pion mass term and the trace anomaly,''
  Phys.\ Rev.\ D {\bf 95}, 016003 (2017)
  [arXiv:1611.04275 [hep-ph]].
  %%CITATION = doi:10.1103/PhysRevD.95.016003;%%
  
\bibitem{Appelquist:2017wcg} 
  T.~Appelquist, J.~Ingoldby, and M.~Piai,
  %``Dilaton EFT Framework For Lattice Data,''
  JHEP {\bf 1707}, 035 (2017)
  [arXiv:1702.04410 [hep-ph]].
  %%CITATION = doi:10.1007/JHEP07(2017)035;%%
  
\bibitem{Gasbarro}
A. Gasbarro, lecture at this conference: %A.~Gasbarro,
  %``Can a Linear Sigma Model Describe Walking Gauge Theories at Low Energies?,''
  EPJ Web Conf.\  {\bf 175}, 08024 (2018)
  [arXiv:1710.08545 [hep-lat]].
  %%CITATION = doi:10.1051/epjconf/201817508024;%%
  
\bibitem{Leino:2017lpc} 
  V.~Leino, J.~Rantaharju, T.~Rantalaiho, K.~Rummukainen, J.~M.~Suorsa, and K.~Tuominen,
  %``The gradient flow running coupling in SU(2) gauge theory with $N_f=8$ fundamental flavors,''
  Phys.\ Rev.\ D {\bf 95}, 114516 (2017)
  [arXiv:1701.04666 [hep-lat]].
  %%CITATION = doi:10.1103/PhysRevD.95.114516;%%
    
\bibitem{Bergner:2016hip} 
  G.~Bergner, P.~Giudice, G.~Münster, I.~Montvay, and S.~Piemonte,
  %``Spectrum and mass anomalous dimension of SU(2) adjoint QCD with two Dirac flavors,''
  Phys.\ Rev.\ D {\bf 96}, 034504 (2017)
  [arXiv:1610.01576 [hep-lat]].
  %%CITATION = doi:10.1103/PhysRevD.96.034504;%%

\bibitem{Rantaharju:2017eej} 
  J.~Rantaharju, C.~Pica, and F.~Sannino,
  %``Ideal Walking Dynamics via a Gauged NJL Model,''
  Phys.\ Rev.\ D {\bf 96}, 014512 (2017)
  [arXiv:1704.03977 [hep-lat]].
  %%CITATION = doi:10.1103/PhysRevD.96.014512;%%

\bibitem{Fodor:2012ty} 
  Z.~Fodor, K.~Holland, J.~Kuti, D.~Nogradi, C.~Schroeder, and C.~H.~Wong,
  %``Can the nearly conformal sextet gauge model hide the Higgs impostor?,''
  Phys.\ Lett.\ B {\bf 718}, 657 (2012)
  [arXiv:1209.0391 [hep-lat]].
  %%CITATION = doi:10.1016/j.physletb.2012.10.079;%%
  
\bibitem{Wong}
C. H. Wong, lecture at this conference:   Z.~Fodor, K.~Holland, J.~Kuti, D.~Nogradi and C.~H.~Wong,
  %``Spectroscopy of the BSM sextet model,''
  EPJ Web Conf.\  {\bf 175}, 08014 (2018)
  [arXiv:1711.05299 [hep-lat]].
  %%CITATION = doi:10.1051/epjconf/201817508014;%%
  
\bibitem{Fodor:2015zna} 
  Z.~Fodor, K.~Holland, J.~Kuti, S.~Mondal, D.~Nogradi, and C.~H.~Wong,
  %``The running coupling of the minimal sextet composite Higgs model,''
  JHEP {\bf 1509}, 039 (2015)
  [arXiv:1506.06599 [hep-lat]].
  %%CITATION = doi:10.1007/JHEP09(2015)039;%%
  
\bibitem{Holland}
K. Holland, lecture at this conference:   Z.~Fodor, K.~Holland, J.~Kuti, D.~Nogradi and C.~H.~Wong,
  %``A new method for the beta function in the chiral symmetry broken phase,''
  EPJ Web Conf.\  {\bf 175}, 08027 (2018)
  [arXiv:1711.04833 [hep-lat]].
  %%CITATION = doi:10.1051/epjconf/201817508027;%%

\bibitem{Fodor:2016wal} 
  Z.~Fodor, K.~Holland, J.~Kuti, S.~Mondal, D.~Nogradi, and C.~H.~Wong,
  %``Electroweak interactions and dark baryons in the sextet BSM model with a composite Higgs particle,''
  Phys.\ Rev.\ D {\bf 94}, 014503 (2016)
  [arXiv:1601.03302 [hep-lat]].
  %%CITATION = doi:10.1103/PhysRevD.94.014503;%%
  
\bibitem{Kuti}
J. Kuti, lecture at this conference: see Ref.~\cite{Fodor}.

\bibitem{Hansen:2017ejh} 
  M.~Hansen, V.~Drach, and C.~Pica,
  %``SU(3) sextet model with Wilson fermions,''
  Phys.\ Rev.\ D {\bf 96}, 034518 (2017)
  [arXiv:1705.11010 [hep-lat]].
  %%CITATION = doi:10.1103/PhysRevD.96.034518;%%
  
\bibitem{Pica}
C. Pica, lecture at this conference: M.~Hansen and C.~Pica,
  %``SU(3) sextet model with Wilson fermions,''
  EPJ Web Conf.\  {\bf 175}, 08018 (2018)
  [arXiv:1710.08801 [hep-lat]].
  %%CITATION = doi:10.1051/epjconf/201817508018;%%

\bibitem{Georgi:1984af}
  H.~Georgi and D.~B.~Kaplan,
  %``Composite Higgs and Custodial SU(2),''
  Phys.\ Lett.\ B {\bf 145}, 216 (1984).
  %%CITATION = doi:10.1016/0370-2693(84)90341-1;%%

\bibitem{Dugan:1984hq}
  M.~J.~Dugan, H.~Georgi, and D.~B.~Kaplan,
  %``Anatomy of a Composite Higgs Model,''
  Nucl.\ Phys.\ B {\bf 254}, 299 (1985).
  %%CITATION = doi:10.1016/0550-3213(85)90221-4;%%
  
\bibitem{Golterman:2015zwa} 
  M.~Golterman and Y.~Shamir,
  %``Top quark induced effective potential in a composite Higgs model,''
  Phys.\ Rev.\ D {\bf 91}, 094506 (2015)
  [arXiv:1502.00390 [hep-ph]].
  %%CITATION = doi:10.1103/PhysRevD.91.094506;%%
  
\bibitem{Golterman:2017vdj} 
  M.~Golterman and Y.~Shamir,
  %``Effective potential in ultraviolet completions for composite Higgs models,''
  Phys.\ Rev.\ D {\bf 97}, no. 9, 095005 (2018)
  [arXiv:1707.06033 [hep-ph]].
  %%CITATION = doi:10.1103/PhysRevD.97.095005;%%

\bibitem{DeGrand:2016htl} 
  T.~A.~DeGrand, M.~Golterman, W.~I.~Jay, E.~T.~Neil, Y.~Shamir, and B.~Svetitsky,
  %``Radiative contribution to the effective potential in composite Higgs models from lattice gauge theory,''
  Phys.\ Rev.\ D {\bf 94}, 054501 (2016)
  [arXiv:1606.02695 [hep-lat]].
  %%CITATION = doi:10.1103/PhysRevD.94.054501;%%
  
\bibitem{Contino:2010rs}
  R.~Contino,
  %``The Higgs as a Composite Nambu--Goldstone Boson,''
  arXiv:1005.4269 [hep-ph].
  %%CITATION = ARXIV:1005.4269;%%

\bibitem{Bellazzini:2014yua}
   B.~Bellazzini, C.~Cs\'aki, and J.~Serra,
   %``Composite Higgses,''
   Eur.\ Phys.\ J.\ C {\bf 74}, 2766 (2014)
   [arXiv:1401.2457 [hep-ph]].
   %%CITATION = ARXIV:1401.2457;%%

\bibitem{Panico:2015jxa}
   G.~Panico and A.~Wulzer,
   %``The Composite Nambu-Goldstone Higgs,''
   Lect.\ Notes Phys.\  {\bf 913}, 1 (2016)
   [arXiv:1506.01961 [hep-ph]].

\bibitem{Ferretti:2013kya} 
  G.~Ferretti and D.~Karateev,
  %``Fermionic UV completions of Composite Higgs models,''
  JHEP {\bf 1403}, 077 (2014)
  [arXiv:1312.5330 [hep-ph]].
  %%CITATION = doi:10.1007/JHEP03(2014)077;%%
  
\bibitem{Kaplan:1991dc}
  D.~B.~Kaplan,
  %``Flavor at SSC energies: A New mechanism for dynamically generated fermion masses,''
  Nucl.\ Phys.\ B {\bf 365}, 259 (1991).
  %%CITATION = NUPHA,B365,259;%%

\bibitem{Ferretti:2014qta} 
  G.~Ferretti,
  %``UV Completions of Partial Compositeness: The Case for a SU(4) Gauge Group,''
  JHEP {\bf 1406}, 142 (2014)
  [arXiv:1404.7137 [hep-ph]].
  %%CITATION = doi:10.1007/JHEP06(2014)142;%%

\bibitem{Will}
W. I. Jay, lecture at this conference.

\bibitem{Venkitesh}
V. Ayyar, lecture at this conference:  
  V.~Ayyar, D.~Hackett, W.~Jay, and E.~Neil,
  %``Confinement study of an SU(4) gauge theory with fermions in multiple representations,''
  EPJ Web Conf.\  {\bf 175}, 08025 (2018)
  [arXiv:1710.03257 [hep-lat]].
  %%CITATION = doi:10.1051/epjconf/201817508025;%%

\bibitem{Dan}
D. Hackett, lecture at this conference: V.~Ayyar, T.~DeGrand, M.~Golterman, D.~C.~Hackett, W.~I.~Jay, E.~T.~Neil, Y.~Shamir, and B.~Svetitsky,
  %``Spectroscopy of SU(4) composite Higgs theory with two distinct fermion representations,''
  Phys.\ Rev.\ D {\bf 97}, no. 7, 074505 (2018)
  [arXiv:1710.00806 [hep-lat]].
  %%CITATION = doi:10.1103/PhysRevD.97.074505;%%

\bibitem{DeGrand:2016mxr} 
  T.~A.~DeGrand, D.~Hackett, W.~I.~Jay, E.~T.~Neil, Y.~Shamir, and B.~Svetitsky,
  %``Towards Partial Compositeness on the Lattice: Baryons with Fermions in Multiple Representations,''
  PoS LATTICE {\bf 2016}, 219 (2016)
  [arXiv:1610.06465 [hep-lat]].
  %%CITATION = ARXIV:1610.06465;%%

\bibitem{Barnard:2013zea} 
  J.~Barnard, T.~Gherghetta, and T.~S.~Ray,
  %``UV descriptions of composite Higgs models without elementary scalars,''
  JHEP {\bf 1402}, 002 (2014)
  [arXiv:1311.6562 [hep-ph]].
  %%CITATION = doi:10.1007/JHEP02(2014)002;%%

\bibitem{Lucini}
B. Lucini and J. W. Lee, lectures at this conference:   E.~Bennett, D.~K.~Hong, J.~W.~Lee, C.-J.~D.~Lin, B.~Lucini, M.~Piai, and D.~Vadacchino,
  %``Higgs compositeness in Sp(2N) gauge theories - Determining the low-energy constants with lattice calculations,''
  EPJ Web Conf.\  {\bf 175}, 08011 (2018)
  [arXiv:1710.06941 [hep-lat]].
  %%CITATION = doi:10.1051/epjconf/201817508011;%%

\bibitem{Vadacchino}
D. Vadacchino, lecture at this conference: E.~Bennett, D.~K.~Hong, J.~W.~Lee, C.-J.~D.~Lin, B.~Lucini, M.~Piai, and D.~Vadacchino,
  %``Higgs compositeness in $\mathrm{Sp}(2N)$ gauge theories --- The pure gauge model,''
  EPJ Web Conf.\  {\bf 175}, 08013 (2018)
  [arXiv:1710.07043 [hep-lat]].
  %%CITATION = doi:10.1051/epjconf/201817508013;%%

\bibitem{Bennett}
E. Bennett, lecture at this conference:   E.~Bennett, D.~K.~Hong, J.~W.~Lee, C.-J.~D.~Lin, B.~Lucini, M.~Piai, and D.~Vadacchino,
  %``Higgs compositeness in $\mathrm{Sp}(2N)$ gauge theories --- Resymplecticisation, scale setting and topology,''
  EPJ Web Conf.\  {\bf 175}, 08012 (2018)
  [arXiv:1710.06715 [hep-lat]].
  %%CITATION = doi:10.1051/epjconf/201817508012;%%

\bibitem{DelDebbio:2017ini} 
    L.~Del Debbio, C.~Englert, and R.~Zwicky,
  %``A UV Complete Compositeness Scenario: LHC Constraints Meet The Lattice,''
  JHEP {\bf 1708}, 142 (2017)
  [arXiv:1703.06064 [hep-ph]].
  %%CITATION = doi:10.1007/JHEP08(2017)142;%%


\end{thebibliography}

%%%%%%%%%%%%%%%%%%%%%%%%%%%%%%%%%%%%%%%%%%%%%%%%%%%%%%%%%%%%%%%%%%%%%%%%%%%%%
\end{document}